# Numerical study of Heat Transfer Enhancement by Deformable Twin Plates in Laminar Heated Channel flow

Rakshitha U. Joshi[1], Atul K. Soti[2] and Rajneesh Bhardwaj[1, 2, *]

[1]Department of Mechanical Engineering.,

[2]IITB-Monash Research Academy,

Indian Institute of Technology Bombay, Mumbai, India 400076

*Corresponding author

Phone: +91 22 2576 7534, fax: +91 22 2576 6875, email: rajneesh.bhardwaj@iitb.ac.in

*Abstract*

Fluid-structure interaction (FSI) of thin flexible fins coupled with convective heat transfer has applications in energy harvesting and in understanding functioning of several biological systems. We numerically investigate FSI of the thin flexible fins involving large-scale flow-induced deformation as a potential heat transfer enhancement technique. An in-house, strongly-coupled fluid-structure interaction (FSI) solver is employed in which flow and structure solvers are based on sharp-interface immersed boundary and finite element method, respectively. We consider twin flexible fins in a heated channel with laminar pulsating cross flow. The vortex ring past the fin sweep higher sources of vorticity generated on the channel walls out into the downstream – promoting the mixing of the fluid. The moving fin assists in convective mixing, augmenting convection in bulk and at the walls; and thereby reducing thermal boundary layer thickness and improving heat transfer at the channel walls. The thermal augmentation is quantified in term of instantaneous Nusselt number at the wall. Results are presented for two limiting cases of thermal conductivity of the fin - an insulated fin and a highly conductive fin. We discuss feasibility of flexible fins and effect of flow cycle-to-cycle variation on the Nusselt number. Finally, we investigate the effect of important problem parameters - Young's Modulus, flow frequency and Prandtl number - on the thermal augmentation.

# 1  Introduction

Thermal augmentation via large-scale flow-induced deformation of flexible fins has engineering applications in piezoelectric fans for thermal management of electronic devices [1] and energy harvesting devices in microsystems [2]. Examples of biological systems include, thermoregulation in elephants via flapping of their large ears [3] and thermal transport enhancement in microchannels using oscillating synthetic cilia [4]. The fluid and structure dynamics with the large-scale flow-induced deformation are coupled with convective heat transfer. The modeling of such problems is computationally challenging and involves moving structure boundaries inside the fluid domain. The coupled flow and heat transfer is highly unsteady with geometric and material non-linearity in structure domain. The coupling of the governing equations of the fluid and structure domain brings additional non-linearity into the system of equations.

Several numerical studies involving large-scale deformation of plates or slender bodies without heat transfer were focused on applications in cardiac flows. Baaijens [5] presented numerical analysis of the FSI of slender bodies placed in a channel flow; however, this study ignored inertial terms in the governing equation of the structure domain. Similarly, Vigmostad et al. [6] showed capability of simulating thin, flexible structures using a sharp-interface Cartesian grid method. Recently studies by Heil et al. [7], Bhardwaj and Mittal [8], Lee and You [9] and Tian et al. [10] validated their respective solvers against the FSI benchmark proposed by Turek and Hron [11]. In this benchmark, an elastic plate attached to the lee side of a rigid cylinder develops self-sustained oscillation in 2D laminar channel flow.

Previous numerical studies showed that the flexible structures can be utilized as a heat transfer enhancement technique, in which the structure was set in the motion by an external source. For instance, Fu and Yang [12] showed that the swinging fins in a heated channel enhances heat transfer which scales with the fin amplitude. Similarly, oscillating fins improves heat transfer from a heat sink due to enhanced convective mixing [13, 14]. Very few numerical studies which considered flow-induced deformation of the structure were reported. For instance, Khanafer et al. [15] simulated a heated flexible cantilever attached to a square cylinder in a channel. However, authors did not investigate thermal augmentation due to the motion of the cantilever. Very recently, Soti and Bhardwaj [16], Shi et al. [17] and Soti et al. [18] demonstrated heat transfer enhancement via large-scale flow-induced deformation in the FSI benchmark proposed by Turek and Hron [11]. These investigations, however, considered the flexible structure along the flow and did not consider configuration in which the structure is mounted in cross flow.

Since utilizing fin is a conventional method to augment heat transfer, it is worthwhile to examine the effectiveness of flexible fins instead of rigid ones. The flexible fins are commonly utilized to improve the heat transfer in nature, for instance, elephants thermoregulate their bodies via flapping of their large ears [3]. In this paper, we perform a proof-of-concept investigation of the effectiveness of flexible fins in cross channel flow harnessing the energy of existing flow field. We employ a state-of-the-art FSI solver (section 2) to investigate the interplay of the flow-field, structure dynamics and convective heat transfer of the deformable elastic fins in a cross-channel flow. Numerical results demonstrating thermal augmentation are discussed in section 3.

## 2 Computational Model

An in-house FSI solver is employed to simulate the fluid and structure dynamics with convective heat transfer. In this solver, a sharp-interface immersed boundary method based on multi-dimensional ghost-cell methodology [19] is utilized and the immersed structure boundary is represented using unstructured grid with triangular elements in Cartesian volume grid of the flow domain. The governing equations of the flow domain are solved on a fixed Cartesian (Eulerian) grid while the movement of the immersed structure surfaces is tracked in Lagrangian framework. The FSI solver employed in the present study was developed by Mittal and co-workers [19-22] and later developed for large-scale flow-induced deformation by Bhardwaj and Mittal [8]. The flow is governed by unsteady, viscous and incompressible Navier-Stokes equations:

$$\frac{\partial v_i}{\partial x_i} = 0 \quad (1)$$

$$\frac{\partial v_i}{\partial t} + \frac{\partial v_i v_j}{\partial x_j} = -\frac{\partial p}{\partial x_i} + \frac{1}{Re}\frac{\partial^2 v_i}{\partial x_j^2} \quad (2)$$

where $i, j$ = 1, 2, 3, and $v_i$, $t$, $p$ and $Re$ are velocity components, time, pressure and Reynolds number, respectively. These equations are discretized in space using a cell-centered, collocated (non-staggered) arrangement of primitive variables $v_i$, $p$ and a second-order, central-difference scheme is used for all spatial derivatives. In addition, face center velocities are computed [23], which results in discrete mass conservation to machine accuracy. The unsteady Navier-Stokes equation is marched in time using a fractional-step scheme [24] which involves two steps: solving an advection-diffusion equation followed by a pressure Poisson equation. During the first step, both the convective and viscous terms are treated implicitly using Crank-Nicolson scheme to improve the numerical stability. In the second step, the pressure Poisson equation is solved with the constraint that the final velocity be divergence-free. Once the pressure is obtained, the velocity field is updated to its final value in the final sub-step. A fast geometric multigrid solver [25] is used to solve the pressure Poisson equation.

The structure dynamics was simulated using an open-source finite-element solver Tahoe$^©$ [26]. The constitutive law for the structure is chosen to be Saint Venant-Kirchhoff material in which the elasticity of the structure is characterized Poisson's ratio and Young's modulus. The structure solver was implicitly (two-way) coupled with the flow solver using a partitioned approach by Bhardwaj and Mittal [8] with numerical stability at low structure-fluid density ratio.

The heat transfer inside the fluid is governed by the following dimensionless energy equation.

$$\frac{\partial T}{\partial t} + v_i \frac{\partial T}{\partial x_i} = \frac{1}{Pe} \frac{\partial^2 T}{\partial x_j^2}, \qquad (3)$$

where $Pe$ is Peclet number and $T$ is dimensionless temperature, defined in terms of dimensional temperature $T^*$, reference wall temperature $T^*_w$ and inlet temperature of the fluid $T^*_{in}$, as follows,

$$T = \frac{T^* - T^*_{in}}{T^*_w - T^*_{in}}, \qquad (4)$$

The heat transfer augmentation is characterized using instantaneous Nusselt number at the channel wall, which is defined as follows [27].

$$Nu(x,t) = \frac{2H}{T_m - 1} \left.\frac{\partial T}{\partial y}\right|_{wall}, \qquad (5)$$

where $2H$ is the dimensionless hydraulic diameter of the channel. The bulk mean temperature, $T_m$, is defined as [27]:

$$T_m(x,t) = \frac{\int_0^H uT dy}{\int_0^H u dy}, \qquad (6)$$

where $u$ stands for dimensionless axial velocity component.

## 2.1 Code validation

The flow solver was extensively validated by Mittal et al. [19] against several benchmark problems such as flow past a circular cylinder, sphere, airfoil and suddenly accelerated circular cylinder and normal plate. The FSI solver was used to simulate and validate 3D biological flows with small scale deformation [20]. The capability of simulating large-scale flow-induced deformation with implicit (two-way) coupling in the FSI solver was implemented and validated by Bhardwaj and Mittal [8]. The validation was carried out against a FSI benchmark proposed by Turek and Hron [11] in which a thin elastic plate attached to a rigid cylinder develops self-induced oscillations in a laminar incompressible channel flow. Very recently, convective heat transfer module including moving immersed boundaries was validated by Soti et al. [18].

# 3 Results and Discussion

We consider twin flexible fins in a 2-D incompressible, laminar flow in a channel of height $2H$. The fins are fixed to the wall, and height and width are $0.8H$ and $0.1H$, respectively (Fig 1A). The geometry is similar to one that was considered by Baaijens [5] to study heart-valve dynamics. We utilize pulsatile velocity at the channel inlet in order to ensure sustained fin motion for thermal augmentation and dissipation of cyclic heat loads, typically needed for localized chip-cooling in microelectronics. The fully-developed inlet velocity is given by,

$$u = 4U\left(\frac{y}{2H}\right)\left(1 - \frac{y}{2H}\right)\left(\frac{1 - \cos(2\pi ft)}{2}\right), \qquad (7)$$

where $U$ is the maximum inlet velocity at the centre of the channel, $2H$ is channel width and $f$ is pulsation frequency. The inlet velocity varies from 0 to $U$ and from $U$ to 0 in first and second half of the cycle, respectively, without flow reversal in full cycle (Fig. 2 (right column)). The boundary conditions for the flow are illustrated in Fig 1A and are described as follows. No-slip is applied at the channel walls and zero Neumann boundary condition is applied for pressure at all boundaries as well as velocity at outlet. Symmetry boundary condition is applied at the top boundary. No-slip and force balance boundary conditions are applied at the fluid-structure interface. The thermal boundary conditions are described as follows. A uniform temperature, $T_0 = 1$, is considered at the channel inlet. The fluid-structure interface is insulated and the channel walls are at constant temperature ($T_w = 1$). At channel outlet zero Neumann temperature boundary condition is applied. Based on thermal conductivity of the fin, two limiting cases - insulated fin and highly conductive fin - are considered. The former corresponds to insulated boundary condition while the latter corresponds to constant temperature ($T = 1$) at the fluid-solid interface. Zero Neumann temperature boundary condition is applied at the outlet. The following problem parameters are selected for the baseline case. The maximum dimensionless velocity at the inlet is $U = 0.5$ and Reynolds number based on fin height and maximum velocity at the inlet is $Re = 150$. The dimensionless Young's modulus, structure-fluid density ratio, Poisson's ratio and Prandtl number are $2 \times 10^3$, 10, 0.4 and 1, respectively.

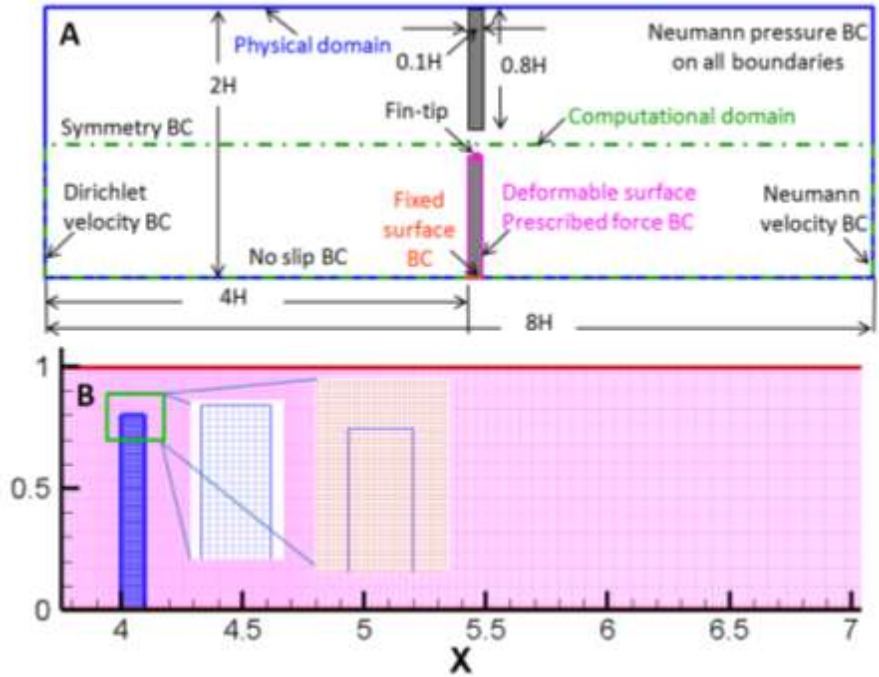

Fig. 1. (A) Schematic with computational domain and boundary conditions (BC). (B) Cartesian mesh and finite element mesh in the computational domain. The zoomed in insets show finite-element mesh in the fin and Cartesian mesh around the fin.

### 3.1 Grid convergence study

We consider three non-uniform grids with $193 \times 97$ (coarse), $257 \times 129$ (medium) and $385 \times 193$ (fine) points in the computational domain to perform grid-convergence study. A typical Cartesian mesh is shown in Fig 1B, with finite element mesh in the fin shown in zoomed-in inset. The minimum grid sizes ($\Delta x_{min} = \Delta y_{min}$) corresponding to these three grids are $1 \times 10^{-2}$, $7.8 \times 10^{-3}$ and $5 \times 10^{-3}$, respectively. Comparison among time history of the fin-tip calculated using the three grids is plotted in Fig 2. The results of the three grids are close and numerical results converge for medium and fine grids, as shown in the insets. The medium grid with 257 × 129 points and $\Delta x_{min} = \Delta y_{min} = 0.0078$ was therefore deemed adequate for resolving the flow field coupled with structure dynamics and we utilized medium grid for all simulations presented in this work.

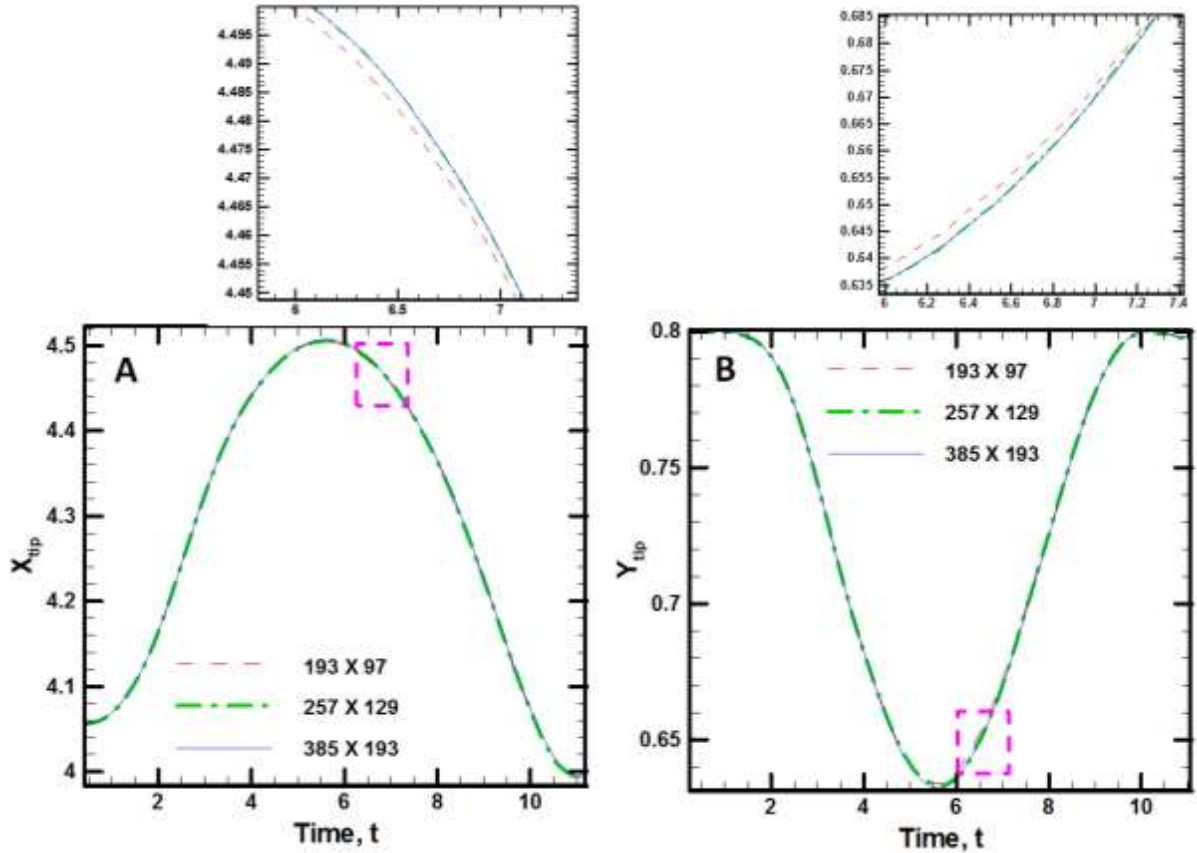

Fig. 2. Grid-convergence study. Three grids are considered with 193 × 97, 257 × 129 and 385 × 193 points in the computational domain. Comparison among the three grids for the time-history of the fin-tip. (A) Tip displacement in X-coordinate. (B) Tip displacement in Y-coordinate. The insets show zoomed-in view of plots marked by magenta dashed box.

### 3.2 Fin dynamics

Fig. 3 shows flow-induced dynamics of the bottom fin during one flow cycle. In Fig. 3 (right column), the fin motion is quantified by plotting time-varying x-coordinate ($X_{tip}$) of the fin-tip, shown as magenta dot on bottom fin in Fig. 1A. We plot magnitude of maximum inlet velocity on symmetry axis ($y = 1$) along with $X_{tip}$ in Fig. 3 (right column). The fin dynamics is mainly governed by the fluid dynamics forces and restoring elastic forces, which are in equilibrium at a given instance. In the first half of the cycle, from $t = 0$ to 5, the inlet velocity as well as fluid dynamics forces increase gradually. These forces are larger near the symmetry axis, owing to the parabolic inlet flow. The fin bends from around $t = 2$ to 7 in a streamlined reconfiguration, resulting in the reduction of frontal area. The maximum deformation of the fin corresponds to its maximum acceleration at around $t = 5$. In the second half of the cycle ($t = 5$ to 10), the restoring elastic forces are dominant as the fluid dynamic forces decrease primarily due to reduction in inlet flow velocity and in the frontal area due to reconfiguration. The fin reaches its initial position with non-zero velocity, which results in a slight backward motion of the fin ($t = 11.5$).

### 3.3 Vortex dynamics

Fig. 3 (left column) shows vorticity contours in the bottom half of the channel during one cycle of the flow. The time instance of a frame in the left column corresponds to black dot shown on the time-varying $X_{tip}$ in the right column. From $t = 4$ to 6, as fins start deforming, rapidly accelerating fluid starts entering in downstream which results in the formation of a wake vortex attached to the fin-tip. This process is analogous to an accelerated flow through an annular orifice [28]. The fluid flowing over the bending fin in the upstream detaches due to boundary layer separation ($t = 4$ to 6) and separated flow forms a wake vortex-ring attached to the fin-tip. The formation of the vortex-ring is attributed to the rolling of the jet by viscous forces of the quiescent fluid in left half of the channel [29]. The growing vortex-ring advances downstream, with the fluid being fed to it by a trailing jet ($t = 6$ to 10). It is attached to the trailing jet while the decelerating fins close as confirmed by Fig. 3 (right column, at $t = 6$ to 10). During this time ($t = 6$ to 10), a positive vortex develops near the channel wall due to back flow, generated by the backward motion of the fin. The vortex ring pinches-off (detaches itself from the trailing jet) at $t = 13$ and moves to downstream, and a new wake vortex starts to form at the fin-tip in next flow cycle ($t = 15$). The vorticity generated on the wall is almost annihilated by the vortex-ring, as evidenced by the residual vorticity footprint downstream. The cross-annihilation of vorticity of a particular sign with the vorticity of opposite sign generated on the wall, results in descending strength of the vortices as they advect along the channel length. Thus, the vortex-ring sweep the higher sources of vorticity generated on the channel walls out into the downstream which aids in the mixing of the fluid in the bulk as well as near the channel walls.

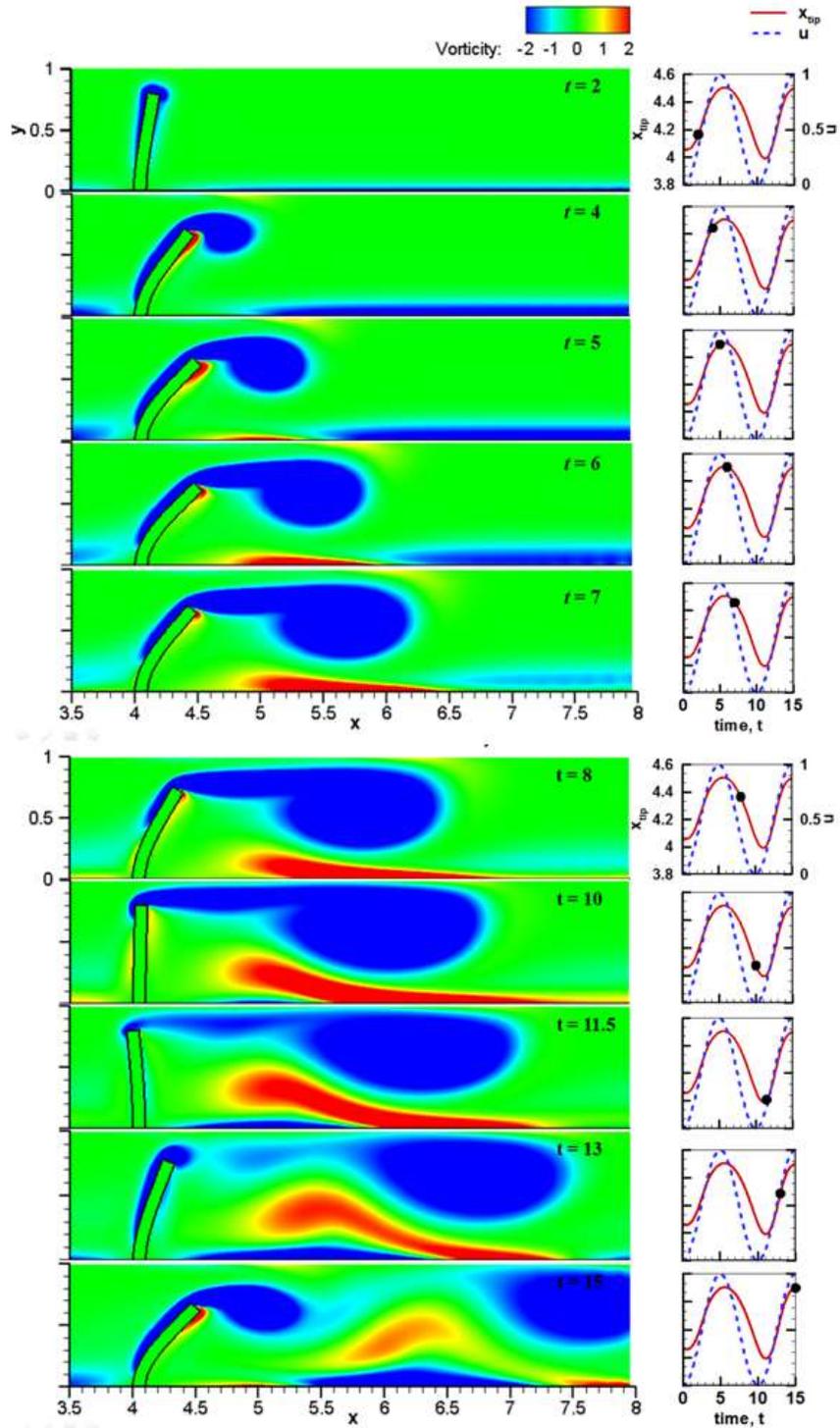

Fig. 3. Vorticity contours in the channel at different time instances, shown in right column as black dots on time-varying x-coordinate of the fin-tip.

## 3.4 Thermal augmentation

We describe thermal augmentation caused by the enhanced fluid mixing described in previous section, in two limiting cases of thermal conductivity of the fin — insulated fin and highly conductive fin. The isotherms are plotted in Fig. 4 in a flow cycle for these two cases in left and right column respectively. The frames correspond to time instances shown as black dots in inset on time-varying $X_{tip}$ during the cycle. The enhanced mixing due to the vortex-ring augments convection in the bulk as well as near the channel walls. The hot fluid near the channel walls moves towards the center of the channel and get replenished by relatively colder fluid, causing thermal enhancement from the walls. The isotherms in Fig. 4 qualitatively confirms thinning of thermal boundary layer at location of the convected vortex, for instance, at $t = 8$ and $x \sim 6$. We note similar characteristics of thermal boundary layer for insulated as well as highly conductive fin in Fig. 4. However, thermal signature for the highly conductive fin shows heat transport from the fin to central region of the channel via the vortex ring ($t = 6$ to $11.5$).

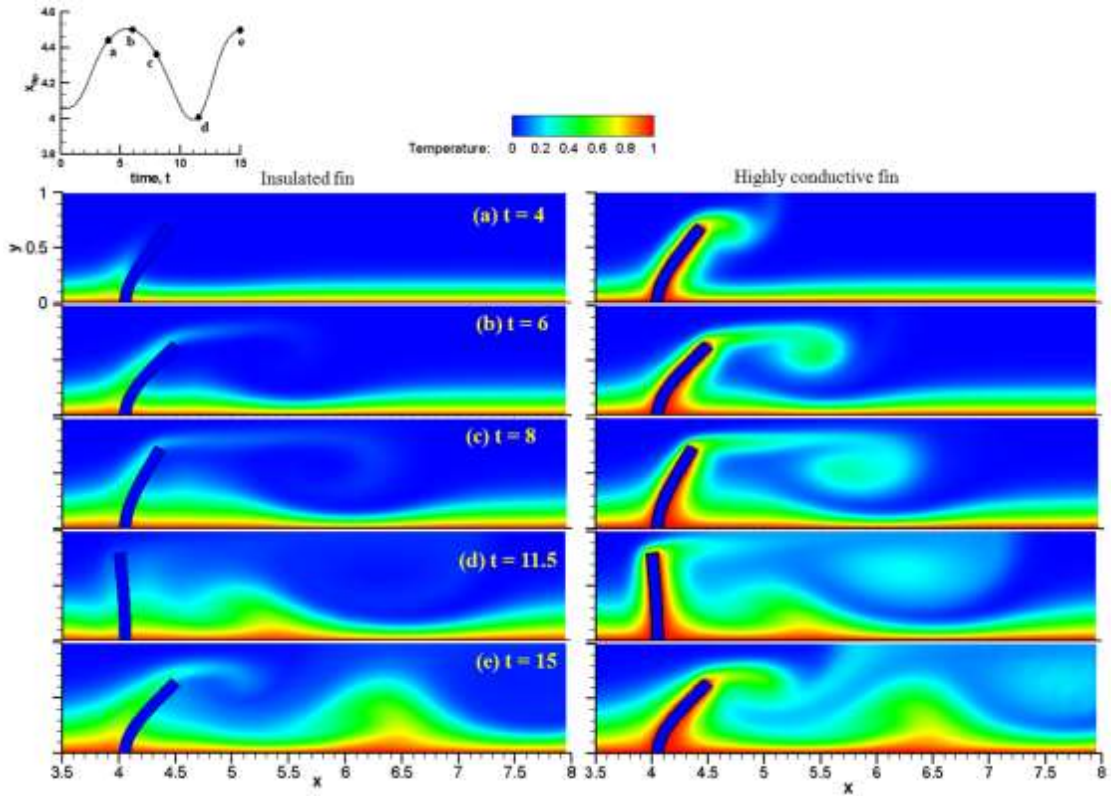

Fig. 4. Isotherms in the channel at different time instances, shown in inset as black dots on time-varying x-coordinate of the fin-tip. Results are shown for the insulated fin and highly conductive fin in left and right column, respectively.

The thermal augmentation is quantitatively quantified by plotting instantaneous Nusselt number ($Nu$) at the bottom wall in Fig 5 for insulated as well as highly conductive fins at five instances, shown as black dots in top-left frame. The peaks of $Nu$ are signature of reduction in thermal boundary layer thickness caused by interaction of the evolving vortex ring and wall vortex. As the vortex ring develops and travels downstream, the peaks accordingly shift downstream for insulated as well as highly conductive fin. For instance, the peak in Nusselt numbers for highly conductive fin shifts from $x = 5.2$ to 6 from $t = 6$ to 8. At $t = 8$, the maximum Nusselt number for highly conductive fin is around 32, 23% larger than that for the insulated fin. This is due to the larger heat transport from the fin to central region of the channel via the vortex ring in former case. We note two peaks for both cases at the instance of pinch-off of the vortex ring ($t = 11.5$). The height of the local peaks decrease due to decrease in strength of the pinched-off vortex as it travels downstream in the channel ($t = 15$). Overall, the augmentation proposed by the present method is suited better for cooling spatially non-uniform heat loads.

### 3.4.1 Feasibility and effectiveness of the flexible fins

As discussed in the previous section, the thermal enhancement obtained by the present method could be effectively employed to extract heat from spatially non-uniform heat sources, for instance, in cooling off hotspots in micoelectronics cooling. Further, we explore possible materials that could be utilized to design such thermal enhancement systems. The dimensionless properties used in the present study are representative of soft elastomers and air. For instance, 10 cm PDMS (polydimethylsiloxane) fins could be used with air with 1-5 cm/s inlet velocity for the parameters given for Case 1 (baseline case) in Table 1. Since typical elastomers has low thermal conductivity (0.1 - 0.5 W/m-K), the thermal augmentation discussed for the insulated fin will be closer to a realistic case as compared to the highly conductive fin.

To verify sustained augmentation in the consecutive cycles., we investigate flow cycle-to-cycle variation of the thermal augmentation for the insulated fin. Fig. 6 compares instantaneous Nusselt number ($Nu$) at the bottom wall at the end of three consecutive cycles, $t = 11, 21$ and $31$. The time-instances are shown as dots on time-varying $X_{tip}$ during the three cycles in the inset. The major changes in $Nu$ are noted at $t = 11$, for $x > 6.5$. The space-variation of Nu is almost similar at $t = 21$ and $31$ and there is no significant difference in the $Nu$ over the last two consecutive cycles. Between first and second cycle, crest and trough in $Nu$ curve appear due to heat due to convection by the pinched-off vortex ring as it passes through $x > 6.5$.

The effectiveness of the present technique is quantified by performing another simulation in only channel without fin. (Case 2, Table 1). Since we get sustained thermal augmentation after first cycle, we compare maximum *Nu* obtained in second cycle. Maximum Nu obtained with fins is 100% larger, confirming the effectiveness of the flexible fins for cooling spatially non-uniform heat loads.

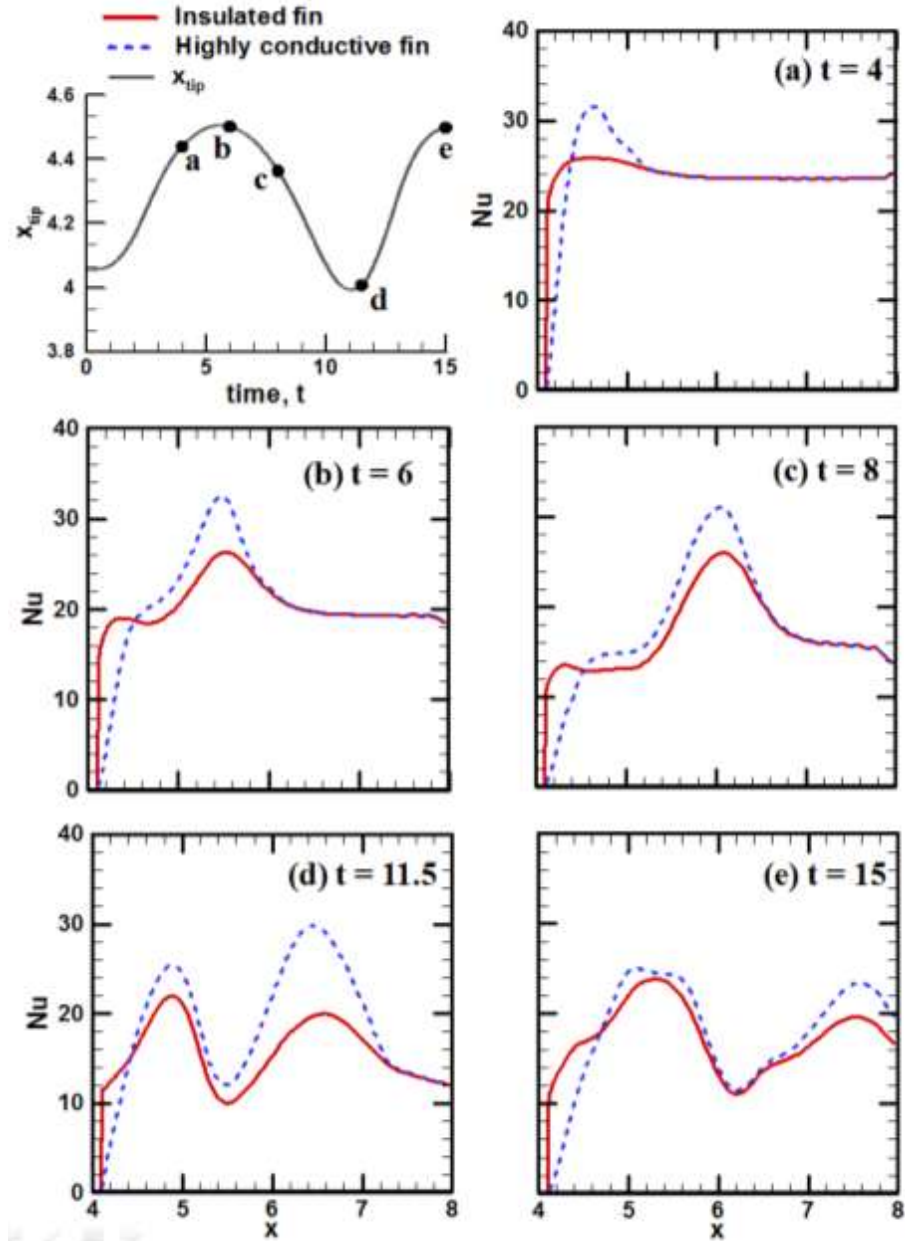

Fig. 5. Instantaneous Nusselt numbers at the bottom channel wall at different time instances which are shown in top-left as black dots on time-varying x-coordinate of the fin-tip. Results are compared for insulated fin and highly conductive fin.

Table 1: Simulation cases performed for the insulated fin.

| Simulation cases | Young's Modulus | Pulsation frequency | Prandtl number | Maximum instantaneous Nusselt number in second cycle |
|---|---|---|---|---|
| Case 1 (baseline) | $2 \times 10^3$ | 0.1 | 1 | 28.2 |
| Case 2 (no fin) | - | 0.1 | 1 | 14.3 |
| Case 3 | $4 \times 10^3$ | 0.1 | 1 | 24.1 |
| Case 4 | $10 \times 10^3$ | 0.1 | 1 | 22.2 |
| Case 5 | $2 \times 10^3$ | 0.05 | 1 | 26.1 |
| Case 6 | $2 \times 10^3$ | 0.2 | 1 | 25.8 |
| Case 7 | $2 \times 10^3$ | 0.1 | 0.1 | 11.2 |
| Case 8 | $2 \times 10^3$ | 0.1 | 10 | 63.3 |

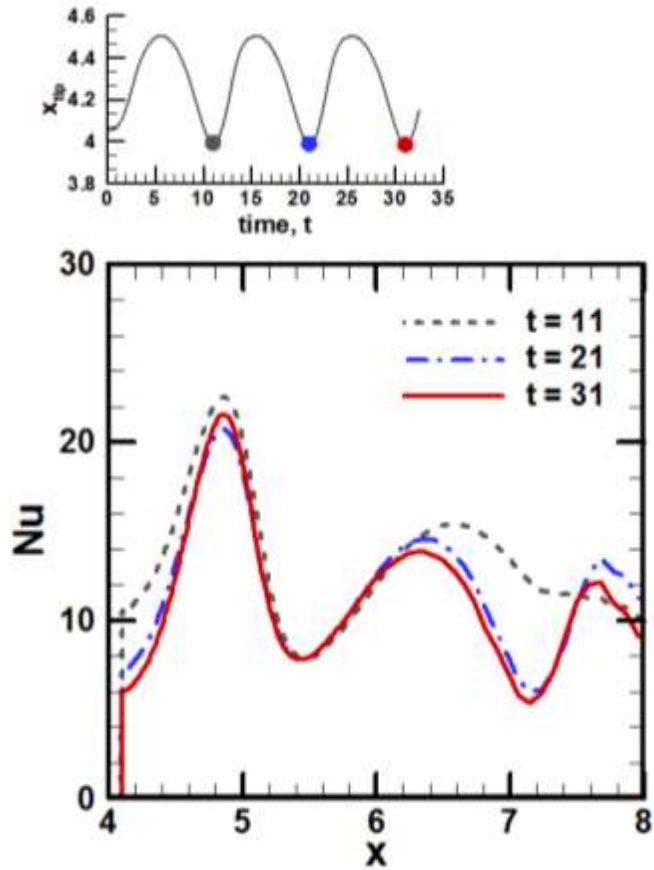

Fig 6: Instantaneous Nusselt number (*Nu*) at the bottom channel wall at different time instances for the insulated fin. *Nu* is compared at end of three cycles at $t = 11$, 21 and 31, shown as dots in the inset on time-varying x-coordinate of the fin-tip.

### 3.4.2 Effect of Young Modulus, pulsation frequency and Prandtl number

In this section, we perform six additional simulations to investigate the effect of Young Modulus (*E*), frequency of the pulsatile inlet flow (*f*) and Prandtl number (*Pr*) on the augmentation. The parameters of the simulation cases and maximum instantaneous Nusselt number in second cycle obtained are summarized in Table 1. First, in the baseline case (Case 1 in Table 1), we vary *E* to $4 \times 10^3$ (Case 3) and $10 \times 10^3$ (Case 4), keeping all other parameters same. As plotted in Fig 7A, the maximum fin deformation inversely scales with *E*. Second, we vary *f* to 0.05 (Case 5) and 0.2 (Case 6). Fig. 7B shows the tip-displacement ($X_{tip}$) for different *f* values and $X_{tip}$ scales with the frequency. Since the fin-dynamics corresponds to the vortex-ring formation and subsequent enhanced heat transfer, the proposed technique is amenable for cyclic heat loads. Finally, *Pr* is varied to 0.1 (Case 7) and 10 (Case 8). In Table 1, *Nu* is the largest for *Pr* = 10 because of larger momentum diffusivity than thermal diffusivity for fluid with larger *Pr*. The isotherms for the three cases of *Pr* are compared at the instance of vortex pinch-off in Fig 8.

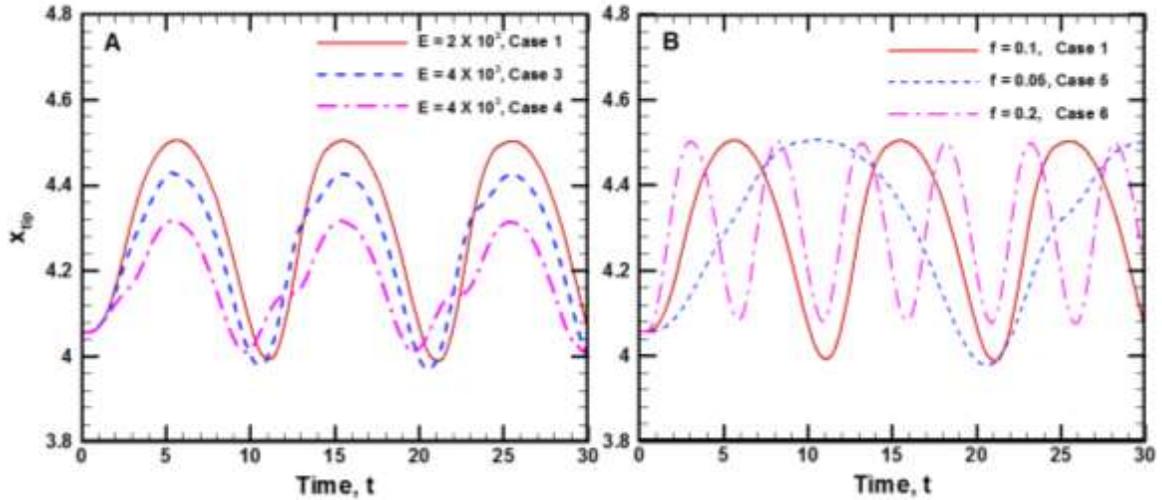

Fig. 7: Effect of Young's Modulus and pulsation frequency. Time-varying x-coordinate of the fin-tip is plotted for different values of for Young's Modulus (A) and pulsation frequency (B).

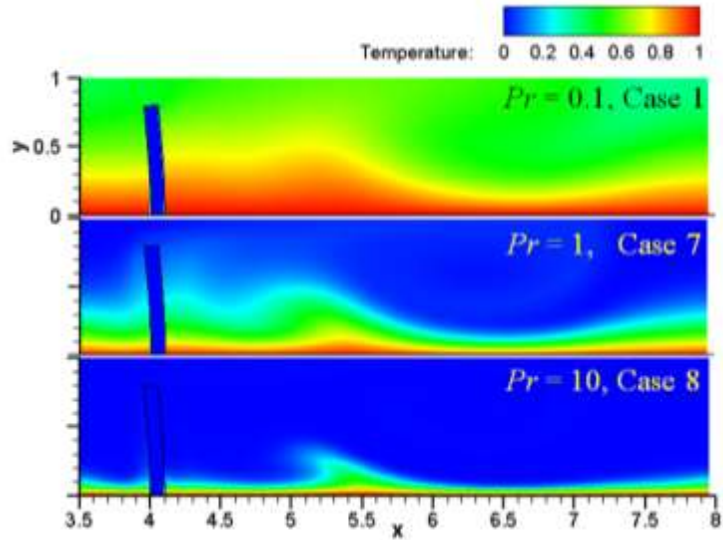

Fig. 8: Isotherms obtained at instance of pinching-off in three cases of Prandtl numbers.

## 4 Conclusions

We investigate vortex dynamics and associated thermal augmentation by twin flexible fins mounted in cross configuration in a heated channel with pulsating inlet velocity. The flow and thermal fields with structure dynamics are solved by an in-house, sharp-interface immersed boundary method based FSI solver. The FSI solver couples a sharp-interface immersed boundary method for flow simulation with a finite-element based structure dynamics solver. An implicit partitioned approach is utilized to ensure the numerical stability of the solver at low structure-fluid density ratio. The vortex ring created due to the flow-induced deformation of the fins, sweeps higher sources of vorticity generated on the channel. Therefore, it enhances fluid mixing and convection in bulk as well as at the walls; reduces thermal boundary layer thickness and enhances heat transfer at the channel walls. The time-varying instantaneous Nusselt number at the wall confirms the thermal augmentation. We discuss feasibility and effectiveness of the flexible fins, and the effect of several parameters — Young's Modulus, pulsation frequency and Prandtl number — on thermal augmentation for conservative case of the insulated fin. The present results provide insights on designing and utilizing flexible fins, potentially useful in dissipating large, non-uniform and cyclic heat loads.

# 5 Acknowledgements

We gratefully acknowledge financial support from Department of Science and Technology (DST), New Delhi, India through fast track scheme for young scientists. R.U.J. was supported partly by Industrial Research and Consultancy Center (IRCC), IIT Bombay through a research internship award for six months.